\documentclass[structabstract]{aa}  
\usepackage{graphicx}
\usepackage{txfonts}
\begin{document}
   \title{Dissecting an intermediate-mass (IM) protostar}
   \subtitle{Chemical differentiation in IC~1396~N}

   \author{A. Fuente
          \inst{1}
          \and
          A. Castro-Carrizo
          \inst{2}
          \and
          T. Alonso-Albi
          \inst{1}
          \and
          M.T. Beltr\'an
          \inst{3}
          \and
          R. Neri
          \inst{2}
          \and
          C. Ceccarelli
          \inst{4}
          \and
          B. Lefloch
          \inst{4} 
          \and
          C. Codella
          \inst{3}
	  \and
          P. Caselli
          \inst{5}
          }
   \institute{ Observatorio Astron\'omico Nacional (OAN), Apdo. 112,
             E-28800 Alcal\'a de Henares, Madrid, Spain
         \and
             Institut de Radioastronomie Millim\'etrique, 300 rue de la Piscine, F-38406 St Martin d'H\`eres, France
          \and
            INAF - Osservatorio Astrofisico di Arcetri, Largo E. Fermi 5, 50125 Firenze, Italy
          \and
           Laboratoire d'Astrophysique Observatoire de Grenoble, BP 53, F-38041 Grenoble C\'edex 9, France
 	\and
          School of Physics \& Astronomy, University of Leeds, Leeds LS2 9JT, UK 
              }
 \abstract 
   {}
   {Our aim is to unveil the physical conditions and structure of the intermediate mass (IM) protostar IRAS~21391+5802 (IC~1396~N) at scales of $\sim$1000~AU. }
   {We have carried out high-angular resolution (1$\farcs$4) observations in the continuum at 3.1mm and in the N$_2$H$^+$ 1$\rightarrow$0, CH$_3$CN 5$_k$$\rightarrow$4$_k$ and $^{13}$CS 2$\rightarrow$1 lines using the Plateau de Bure Interferometer (PdBI). In addition, we have merged the PdBI images with previous BIMA (continuum data at 1.2mm and 3.1mm) and single-dish (N$_2$H$^+$ 1$\rightarrow$0) data to have a comprehensive description of the region.}
   {The combination of our data with BIMA and 30m data show that the bipolar outflow associated has completely eroded the initial molecular globule. The 1.2mm and 3.1mm continuum emissions are extended along the outflow axis tracing the warm walls of the biconical cavity. Most of the molecular gas, however, is located in an elongated feature in the direction perpendicular to the outflow. 
A strong chemical differentiation is detected across the molecular toroid, with the N$_2$H$^+$ 1$\rightarrow$0 emission absent in the inner region.
}
 {Our PdBI data show two different regions in IC1396~N: (i) the young stellar objects (YSO) BIMA 3 and the protocluster BIMA 2, detected in dust continuum emission and one of the individual cores, IRAM 2A, in the CH$_3$CN 5$_k$$\rightarrow$4$_k$ line, and (ii) the clumps and filaments that have only been detected in the N$_2$H$^+$ 1$\rightarrow$0 line. The clumps belonging to this second group are located in the molecular toroid perpendicular to the outflow, mainly along the walls of the biconical cavity. This chemical differentiation can be understood in terms of the different gas kinetic temperature. The [CH$_3$CN]/[N$_2$H$^+$] ratio increases by 5 orders of magnitude with gas temperature, for temperatures between 20~K and 100~K. The CH$_3$CN abundance towards IRAM~2A is similar to that found in hot corinos and lower than that expected towards IM and high mass hot cores. This could indicate that IRAM~2A is a low mass or at most Herbig Ae star (IRAM 2A) instead of the precursor of a massive Be star. Alternatively, the low CH$_3$CN abundance could also be the consequence of IRAM 2A being a Class 0/I transition object which has already formed a small photodissociation region (PDR).}
   \maketitle
%

\section{Introduction}

Intermediate-mass young stellar objects (IMs) (protostars and Herbig
Ae/Be stars with M$_\star$ $\sim$2$-$10\,M$_{\odot}$) are crucial to
star formation studies because they provide the link between the
evolutionary scenarios of low- and high-mass stars.  These objects
share many similarities with high-mass stars, in particular they
are predisposed to be formed in clusters.
However, to study them presents decided advantages compared to
massive star forming regions, as many of them are located close to the
Sun ($\leq$1\,kpc) and in regions of reduced complexity.

IRAS~21391+5802 (IC~1396~N) is one of the best studied
IM protostars (L=440\,L$_\odot$, d=750\,pc). 
Classified as a Class 0\,/\,I
borderline source, this young protostar is associated with a very
energetic bipolar outflow. In addition,
near-infrared images by Nisini et al.\ (2001) and Beltr\'an et al. (2009)
revealed the presence of a collimated 2.12~$\mu$m H$_2$ jet. 
The outflows and envelope of this protostar were firstly 
mapped by Codella et al. (2001) and Beltr\'an et al. (2002, 2004b) using BIMA and OVRO. 
These observations revealed the
existence of 3 millimeter continuum sources in the region. The most intense
one, BIMA~2, is located at the center of the envelope and seems to be the driving source
of the most energetic outflow.

Recently, Neri et al. (2007) reported high angular resolution continuum images 
at 3mm and 1.3mm carried out with the IRAM Plateau de Bure Interferometer (PdBI) at its
most extended configuration. 
The high sensitivity and spatial resolution of the two
continuum images clearly testify to the presence of at least three bright
continuum emission cores at the position of the source previously named BIMA 2. 
While the two weaker cores were
not resolved by the interferometer, the primary core IRAM 2A was
resolved at 1.3mm emission to an elliptical region of
$\sim$300\,AU$\times 150$\,AU. The mass and dust emissivity spectral index
of this core are similar to those measured in circumstellar disks
around Herbig Ae/Be stars (Neri et al. 2007; Fuente et al. 2003, 2006; 
Alonso-Albi et al. 2008, 2009a). 
Other possible interpretations (hot corinos,
cold compact pre-stellar clumps) cannot be, however, discarded.

In this Paper, we present high angular resolution images of the N$_2$H$^+$ 1$\rightarrow$0,
CH$_3$CN 5$_k$$\rightarrow$4$_k$ and $^{13}$CS J=2$\rightarrow$1 lines observed with the PdBI 
in its extended AB configuration. In addition we combine our previous 
continuum PdBI images (Neri et al.2007) with the BIMA observations published by Beltr\'an et al. (2002).
The new 1.2mm and 3.1mm continuum images together with the molecular 
data provide a valuable insight into the chemical and physical structure of this IM protocluster, in 
particular of the IM hot core.

\section{Observations}

\subsection{PdBI observations}

Observations were performed with the PdBI in
AB configuration between January and March 2008. We observed a 1.4~GHz
bandwidth with receivers tuned at 91.78~GHz. The spectral
configuration allowed achieving a resolution of 0.25~km/s for the
transitions CH$_3$CN 5$_k$$\rightarrow$4$_k$, N$_2$H$^+$ 1$\rightarrow$0 and 
for $^{13}$CS 2$\rightarrow$1. Since the lines are wide, all the maps have
been created with a spectral resolution of 0.2 MHz ($\sim$0.64~km/s). 
Continuum was obtained by
averaging the observed band with no line contribution. Because of some
poor quality data, the uv-coverage is
slightly different for the different lines. Synthesized beams
are 1.56$"$$\times$1.23$"$ PA 68$^\circ$ for CH$_3$CN, 1.38$"$$\times$1.16$"$ PA 95$^\circ$
for N$_2$H$^+$ and 1.42$"$$\times$1.21$"$ PA 98$^\circ$ for $^{13}$CS. The rms
of the images is $\sim$ 2.5 mJy/beam (0.2 K).

In order to calibrate the changes in time of the complex gains
(i.e. phases and amplitudes), we observed alternatively with the
source, every 22min, 2037+511. The calibration was found to be
straightforward for all the tracks. A flux of 1.2~Jy was deduced for
2037+511 at the observed frequency. The RF calibration was performed
by observing for few minutes the brightest sources in the sky. MWC349
was observed and used as flux reference, adopting a flux of 1.1~Jy as
suggested by recent PdBI models and consistently with expected
antennas' efficiency values. The flux calibration is found to be
reliable to better than 10\%.

 \subsection {PdBI+BIMA Continuum images}

The PdBI 3mm continuum data are merged with
previous PdBI data by Neri et al (2007), and with older interferometric
BIMA observations by Beltr\'an et al (2002). The contrast of weights of
the different observations makes it difficult the merging, forcing to
lower the weights of the PdBI data.

At 1mm we merged the PdBI data by Neri et al (2007) and the BIMA
observations by Beltr\'an et al (2002). The merging was corrected by the
different beam sizes, no factor was added to correct for weight
contrast. The merged image is only reliable for the innermost
25 arcseconds, which corresponds to $\sim$HPBW of the PdBI primary beam.
Due to the PdBI primary beam correction, the imaging becomes very noisy
close to the border of the PdBI primary beam. Despite nearly a factor of 2.5
difference in the primary beam sizes, we are confident in the elongated structure 
detected within the PdBI
primary beam along the outflow direction.

Finally, note that the data fluxes were not corrected for the
different frequencies of the data sets; the BIMA data frequencies are
97.98 and 244.94 GHz, those of the previous PdBI observations are
92.09 and 242.00 GHz, and last 3mm continuum data center at 91.78
GHz. Assuming the largest spectral index measured in the region,
alpha=+2.8 (Neri et al. 2007), we could introduce a maximum error in
the derived continuum fluxes of ~ 30\% at 3mm and of ~ 10\% at 1mm,
which applies mainly to the cocoon.

\subsection{N$_2$H$^+$ 1$\rightarrow$0 image}
The PdBI N$_2$H$^+$ data were merged with short-spacing
observations obtained with the 30m-telescope (at Pico Veleta,
Spain). A 120"-size field was mapped with the single dish, by
observing every ~ 12".
As commented in Sect 3.3, the flux is mostly filtered out in the
interferometric data, for that the single-dish data are so crucial to recover
the flux and to reconstruct the line emission distribution.

\section{Small scale ($\leq$~1000~AU)}

\subsection{CH$_3$CN: The hot core IRAM 2A}
The CH$_3$CN 5$_k$$\rightarrow$4$_k$ line emission has only been
detected towards IRAM 2A, the most massive core in the
proto-cluster. The integrated intensity map
of the CH$_3$CN 5$_k$$\rightarrow$4$_k$ transition shows that the
emission arises in a point source centered at the position of
the massive hot core IRAM 2A (see Fig. 1).

Comparing the integrated emission peak of the PdBI map with that of the 
30m spectrum (see Fig. 2), we estimate that $\sim$1--2\% of the emission is coming 
from a compact region around IRAM 2A while the remaining
arises from an extended component. This is approximately the same
number we obtain by dividing the 1.3mm flux towards IRAM 2A, 35~mJy (Neri et al.
2007) and the single-dish 1.3mm flux, 1.4~Jy. This shows that CH$_3$CN is 
a good tracer of IM hot cores.

The interferometric profile of the CH$_3$CN 5$_k$$\rightarrow$4$_k$ emission
is very different from that observed with the 30m telescope (see Fig. 2). 
In fact, there is a shift of $\sim$2.5~km/s between the centroid
of the PdBI emission and that of the 30m. However, it is much more similar to
that observed in the high excitation CH$_3$CN 14$_k$$\rightarrow$13$_k$ lines which suggests that the difference
between the single-dish and interferometric profiles of the CH$_3$CN 5$_k$$\rightarrow$4$_k$
is mainly due to the very different angular resolution of the observations (the beam of the 30m telescope at the
frequency of the CH$_3$CN 5$_k$$\rightarrow$4$_k$ line is $\sim$27$"$, the 30m beam at the frequency of the
CH$_3$CN 14$_k$$\rightarrow$13$_k$ is $\sim$9$"$, in our PdBI observations the beam is 1.56$"$$\times$1.23$"$). 
It suggests that the velocity of the hot core is different from that of the bulk of the
molecular cloud. Most of the CH$_3$CN 5$_k$$\rightarrow$4$_k$ single-dish emission is arising in the extended envelope 
whose kinematics has been severely affected by the bipolar outflow (see Codella et al. 2001,  Beltr\'an et al. 2002, 2004b).
However, the different velocity profiles of the PdBI and 30m lines could also be due to the 
filtering of the extended emission by the PdBI observations. The amount of missed flux is different for each spectral 
channel depending on the spatial distribution of the emission at that velocity.
Interferometric observations at lower angular resolution are required to 
discern about filtering effects and a real change in velocity between the hot core and the surrounding cloud. 

We have determined the CH$_3$CN column density towards IRAM 2A using the rotational diagram technique
and the interferometric 13$_k$$\rightarrow$12$_k$ data published by Neri et al. (2007), those presented in this Paper and
the CH$_3$CN 14$_k$$\rightarrow$13$_k$ line observed with the 30m telescope (see Fig. 3). We assume that all the
emission of the CH$_3$CN 14$_k$$\rightarrow$13$_k$ line arises in the hot core IRAM~2A. 
In the case that an extended emission component is also contributing to the flux of the CH$_3$CN 14$_k$$\rightarrow$13$_k$ line, 
the derived temperature would be an upper limit to the actual one.
We derive a rotational temperature of 97$\pm$25~K and a CH$_3$CN column density of 6.5$\pm$5.0 10$^{13}$~cm$^{-2}$
averaged in a beam of 1.6$"$$\times$1.2$"$. Assuming that the dust temperature is 100~K, and the 
size of IRAM 2A estimated by Neri et al (2007), 0.4$"$$\times$0.2$"$, we obtain
a CH$_3$CN abundance of 5$\pm$3~10$^{-10}$ in this core.

\begin{figure}
\includegraphics{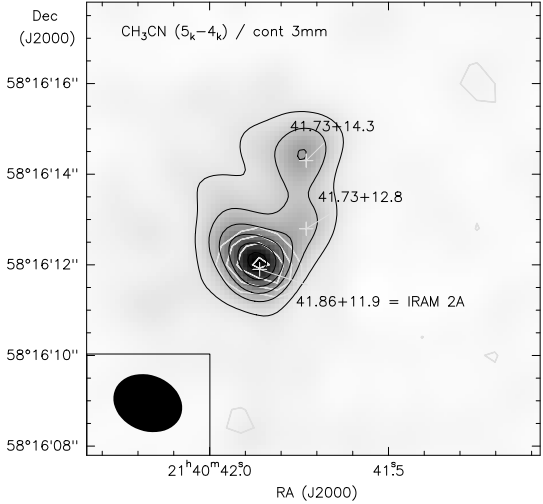}
\vspace{7.5cm}
      \caption{Integrated intensity maps of the CH$_3$CN 5$_k$$\rightarrow$4$_k$ (white contours)
superposed on the 3mm continuum image (grey scale map). Contours for the CH$_3$CN image are  
34~mJy/beam$\times$km/s (5$\times$$\sigma$) to 75~mJy/beam$\times$km/s by 9~mJy/beam$\times$km/s. 
Contours for the 3mm continuum image are 1~mJy/beam to 6~mJy/beam by steps of 1~mJy/beam.}
         \label{Fig 3}
 \end{figure}

\begin{figure}
\includegraphics{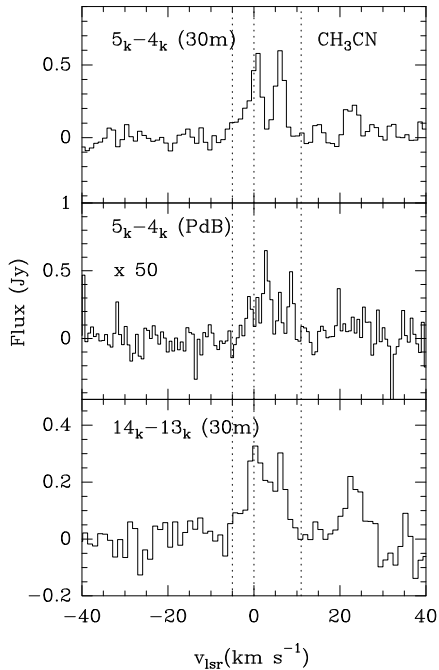}
\vspace{9.0cm}
      \caption{Spectra of the CH$_3$CN 5$_k$$\rightarrow$4$_k$ and CH$_3$CN 14$_k$$\rightarrow$13$_k$ observed 
with the 30m and PdBI towards IRAM 2A. Dashed lines indicate the ambient velocity, v$_{lsr}$=0~km~s$^{-1}$, and the
velocities v$_{lsr}$=-5 and 11~km~s$^{-1}$ for an easier comparison between the 30m and PdBI spectra. }
         \label{Fig 2}
 \end{figure}

\begin{figure}
\includegraphics{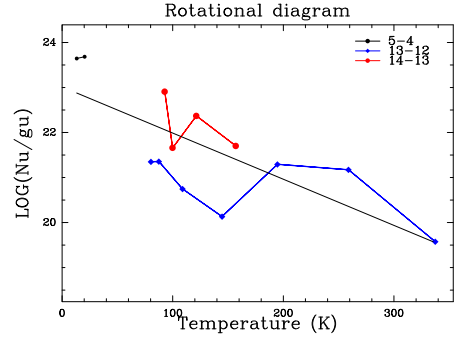}
\vspace{5cm}
      \caption{Rotational diagram of CH$_3$CN towards IRAM 2A. The rotational diagram has been built using the 
interferometric 13$_k$$\rightarrow$12$_k$ data published by Neri et al. (2007) (blue), those presented in this Paper (black) and
the CH$_3$CN 14$_k$$\rightarrow$13$_k$ line as observed with the 30m telescope (red).
}
         \label{Fig 3}
 \end{figure}

\subsection{$^{13}$CS}
We have not detected the $^{13}$CS 2$\rightarrow$1 line in this region. Beltr\'an et al. (2004b) mapped the region in
the CS J=2$\rightarrow$1 line using BIMA with an angular resolution of $\sim$7.0$"$$\times$6.3$"$. The most intense peak is detected towards BIMA 2 with a maximum intensity of 9.6~Jy beam$^{-1}$ km s$^{-1}$ integrated in a velocity interval of -3 to 3 km/s. Assuming that the line ratio between the $^{12}$CS 2$\rightarrow$1/$^{13}$CS 2$\rightarrow$1 is $\sim$60 (a reasonable value for N(CS)=2~10$^{14}$~cm$^{-2}$ derived by Codella et al. 2001), our upper limit to the $^{13}$CS emission implies that less than 60 \% of the CS emission detected by BIMA interferometer is arising in the IM hot core.
In the case of a smaller $^{12}$CS 2$\rightarrow$1/$^{13}$CS 2$\rightarrow$1  line ratio, i.e., a higher opacity of the
CS 2$\rightarrow$1 line, the fraction of the $^{13}$CS emission arising in the hot core would be smaller.
Thus, the $^{13}$CS 2$\rightarrow$1 line does not seem to be a good tracer of IM hot cores.
This is consistent with the kinematical study by Beltr\'an et al. (2004b). They concluded that the CS emission is not
tracing the dense hot core, but the interaction of the molecular outflow with the core(s).

\subsection{N$_2$H$^+$}
In Fig. 4 we show the integrated intensity maps of the N$_2$H$^+$ J=1$\rightarrow$0 F=1$\rightarrow$1, F=2$\rightarrow$1 and F=0$\rightarrow$1 lines. Emission of the N$_2$H$^+$ line has been detected in several clumps across the area sampled by the
interferometer (labeled $N$ in Fig 4 ), but none of them is spatially coincident with the compact cores detected in the continuum images (see Fig. 5). Moreover, the interferometric spectrum towards the most intense core, IRAM 2A, has no evidence  of emission. We have convolved the whole image with a beam of 27$"$ to compare the result with the single-dish spectrum obtained with the 30m  by Alonso-Albi et al. (2009).
We find that only 1\% of the flux was recovered by the interferometer. The velocity profile of the PdBI and 30m
spectra are also quite different showing that most of the emission at ambient and red-shifted velocities has been resolved and filtered out by
the PdBI observations (see Fig. 6). 

In Fig 7, we show our N$_2$H$^+$ 1$\rightarrow$0 F=2$\rightarrow$1 image superposed to the bipolar molecular outflow as traced by
the interferometric observations of CO 1$\rightarrow$0 and CS 5$\rightarrow$4 lines. All the N$_2$H$^+$ clumps except N5 seem to
follow the morphology of the bipolar outflow, delineating the walls of the cavity. The fact that filtering is less important at velocities shifted from that of the ambient cloud, favors the detection of the interaction layer between the outflow and the surrounding cloud. 
The match between N$_2$H$^+$ and the bipolar outflow lobes is better when we compare with the CS 5$\rightarrow$4 line. This is the expected behavior since the N$_2$H$^+$ and CS are high dipole moment molecules that trace the dense gas. In fact, CS and
N$_2$H$^+$ could have similar spatial distribution in the blue lobe if we take into account the lower angular resolution of the CS data. 

\begin{figure*}
\includegraphics{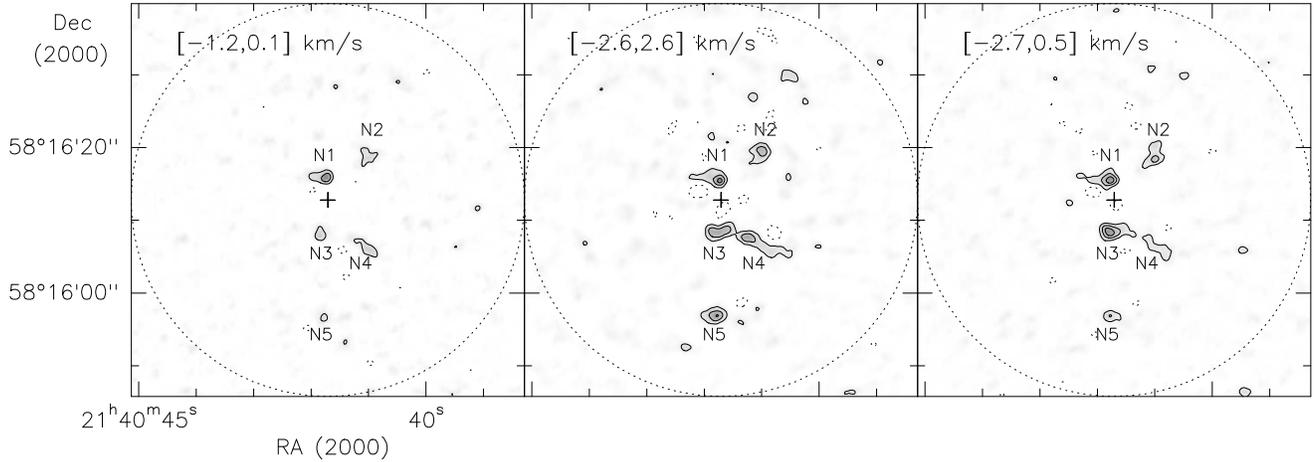}
\vspace{7cm}
      \caption{Integrated intensity images of the N$_2$H$^+$ J=1$\rightarrow$0 F=1$\rightarrow$1 ({\it left}), F=2$\rightarrow$1 ({\it middle})
and  F=0$\rightarrow$1 ({\it right}) lines. First contour and contour spacing is 3$\times$$\sigma$, where $\sigma$ is
4.2mJy/beam({\it left}), 9.1mJy/beam ({\it middle}), and 6.0 mJy/beam ({\it right}). The primary beam ($\approx$54$"$) is indicated
by a dotted circle.}
         \label{Fig 6}
 \end{figure*}

\begin{figure}
\includegraphics{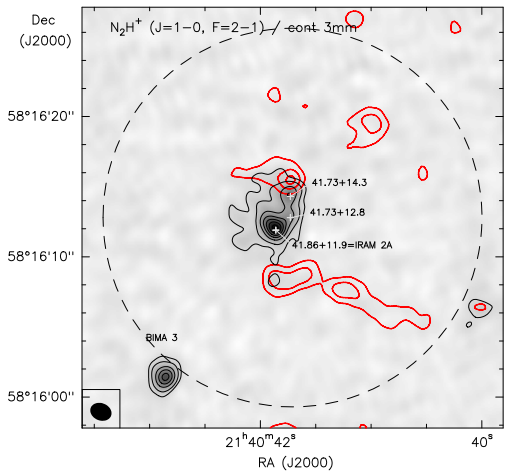}
\vspace{7.5cm}
      \caption{Integrated intensity map of the N$_2$H$^+$ 1$\rightarrow$0 F=2$\rightarrow$1 line (red contours)
superposed on the 3mm continuum image (grey scale map). Contours of the N$_2$H$^+$ 1$\rightarrow$0 F=2$\rightarrow$1 line
are the same as in Fig. 4. Contours of the 3.1mm continuum emission are 0.5~mJy/beam and 1~mJy/beam to 6~mJy/beam by steps of 1~mJy/beam. The primary beam is indicated by a dashed line.}
         \label{Fig 5}
 \end{figure}

\begin{figure}
\includegraphics{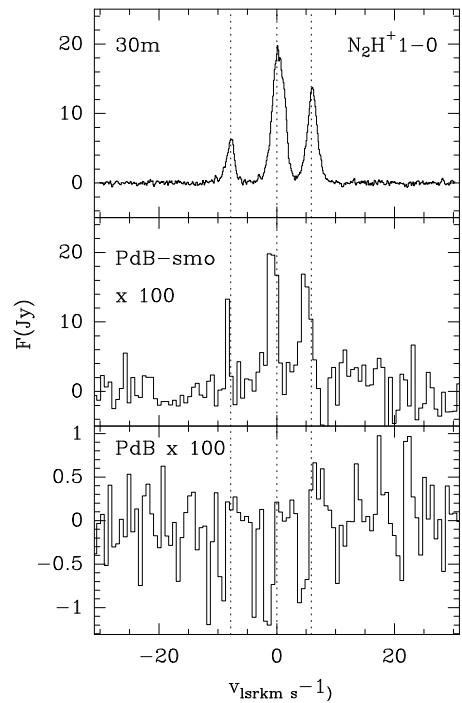}
\vspace{9.5cm}
      \caption{The spectrum of the N$_2$H$^+$ 1$\rightarrow$0 line obtained from our interferometric images towards IRAM 2A is shown in the bottom panel. In the middle panel, we show the spectrum obtained after convolving the interferometric N$_2$H$^+$ 1$\rightarrow$0 image with a beam of 27$"$. In the upper panel we show the spectrum of the same line obtained within the 30m telescope. Dashed lines indicate the position of the F=1$\rightarrow$1, F=2$\rightarrow$1 and F=0$\rightarrow$1 hyperfine components.}
         \label{Fig 4}
 \end{figure}

\begin{figure}
\includegraphics{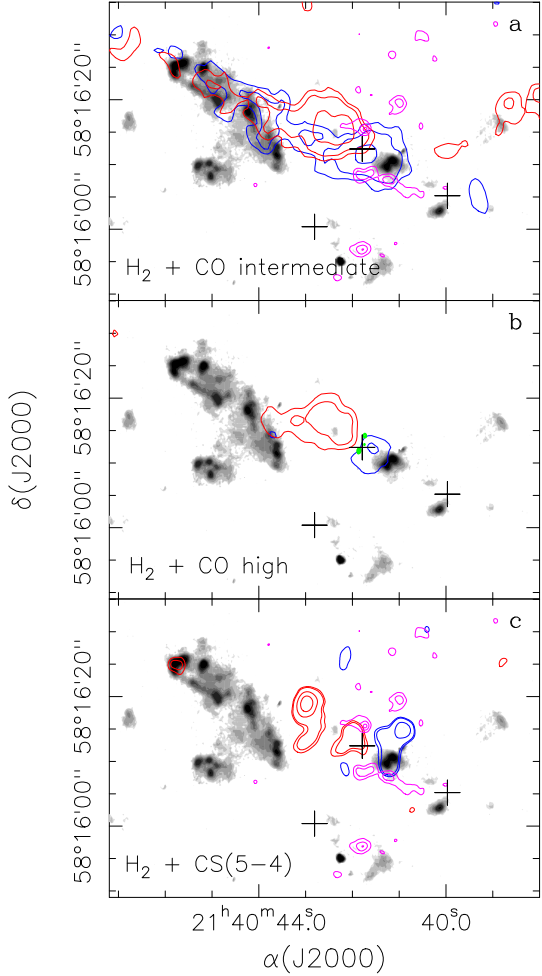}
\vspace{13.5cm}
      \caption{Continuum 1.2mm sources IRAM 2A and 41.73+12.8 (green contours in the middle panel) and contours of the integrated intensity of the N$_2$H$^+$ 1$\rightarrow$0 F=2$\rightarrow$1 line (pink contours in the upper and bottom panels) superposed on the interferometric maps of the CO outflow published by Beltr\'an et al. (2001). Contours of the 1.2mm continuum emission are 2.5, 3.0 to 21.0 by 3.0 mJy/beam. For the 
N$_2$H$^+$ integrated intensity emission, the contours are 27.3, 54.6, 81.9 mJy/beam$\times$km~s$^{-1}$ 
The grey scale is the 2.12 $\mu$m H$_2$ emission from Beltr\'an et al. (2009). Crosses indicate the positions of BIMA~1,
BIMA~2 and BIMA~3 }
         \label{Fig 5}
 \end{figure}

\section{Large Scale}

\subsection{PdBI+BIMA images continuum images}

 The continuum images at 1.2mm and 3.1mm resulting from merging BIMA and PdBI data show two different emission components: (i) the central cluster identified by Neri et al. (2007) and (ii) extended emission along the outflow direction (see Fig. 8).  The emission from the central cluster was already modeled by 
 Neri et al. (2007). They found that it is well explained as arising from three compact clumps 
 ($<$ 300~AU) immersed in a cocoon of about $\sim$2800~AU. The spectral indices are different for the cocoon and
 the compact cores. While the compact cores have spectral indices of $\sim$1.4--1.9, the spectral index in the
 cocoon is $\sim$2.8. The value in the cocoon is consistent with optically thin dust emission with $\beta$$\sim$1.

 The continuum emission at 1.2mm and 3.1mm is extended along the outflow axis. This extended emission arises from the walls of the cavity excavated by the outflow, that are expected to be warmer than the surroundings. The dust continuum emission tracing the walls excavated by the outflow has also been observed in low-mass star-forming regions, as for example in the Class 0 L1157 (Beltr\'an et al. 2004a). 
The spatial distribution of the continuum emission at 1.2mm is different from that at 3.1mm. The 1.2mm emission is
more intense in the eastern lobe. In contrast, the 3.1mm emission comes mainly from the western lobe. 
This different spatial distribution could reveal a gradient in the 1.2mm/3.1mm spectral index along the outflow.
While values larger than 2, consistent with optically thin dust emission, are found in the eastern (red) lobe,
values $\sim$0.6, typical for an ionized wind, are found in the western (blue) lobe. 
However, taking into account the technical complexity and the uncertainties involved in the merging processes of the PdBI and BIMA data,
we have to be cautious with this result. Moreover, in regions with complex morphologies, the different filtering of the continuum emission at 3.1mm and 1.2mm (different synthesized beams) can produce an apparent change in the 1.2mm/3.1mm spectral index. A more complete set of data with visibilities that provide uv-coverage with critical sampling and
consistent frequency scaling is required to confidently measure the 1.2mm/3.1mm spectral index. 

\begin{figure}
\includegraphics{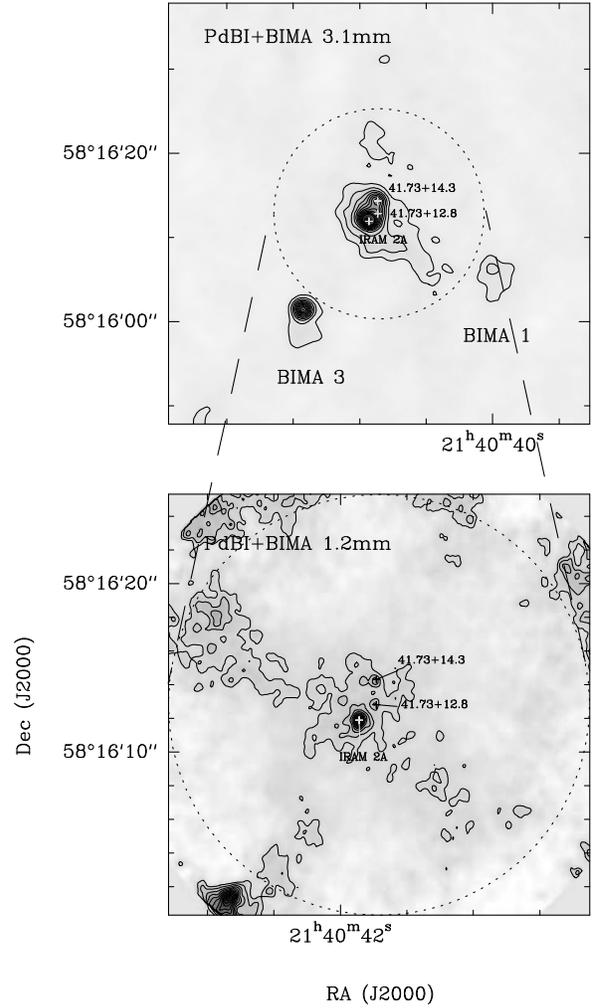}
\vspace{13.7cm}
      \caption{PdBI+BIMA continuum images at 3.1mm (upper panel) and 1.2mm (bottom panel). Contours are: 0.5 mJy/beam to 6 mJy/beam by steps of 0.5 mJy/beam in the 3.1mm image, and 3.5 mJy/beam to 39.5 mJy/beam by steps of 3.0 mJy/beam in the 1.2mm continuum image. The primary beam of the PdBI at 1.2mm is drawn in the panels. 
}
         \label{Fig 1}
 \end{figure}


\subsection{30m+PdBI N$_2$H$^+$ 1$\rightarrow$0 image}

We have included the 30m short spacing data 
and synthesized the 30m+PdBI image to have a more realistic view of the spatial distribution
of N$_2$H$^+$ emission. After merging the 30m and PdBI data we have obtained the image shown in Fig. 9.
The highest red and blue shifted velocities show the same spatial distribution as the clumps N  
in the PdBI image (see 3.65 km s$^{-1}$ and -1.38 km s$^{-1}$ panels in Fig. 9). 
This high velocity gas is distributed in two filaments in the North and South of IRAM 2A respectively.
The bulk of the N$_2$H$^+$ emission is, however, arising in an elongated envelope located perpendicular to the molecular outflow (see Fig. 7 and 8). This envelope does present a velocity gradient of $\sim$1~km~s$^{-1}$ over an
angular distance of 11$"$ ($\sim$ 23 km s$^{-1}$ pc$^{-1}$) in the direction of the bipolar outflow. This velocity gradient is consistent with the outflow kinematics which corroborates that the kinematics of the whole molecular globule is affected by the bipolar outflow.

The N$_2$H$^+$ envelope is elongated in the same orientation 
as the cocoon in the continuum model proposed by Neri et al. (2007). This suggests that both emissions are tracing the same physical structure
and that the different spatial distribution of the emission is due to chemical  differentiation. As discussed in Sect. 5, the [CH$_3$CN]/[N$_2$H$^+$] ratio is strongly dependent on the gas and dust temperature.

\begin{figure*}
\includegraphics{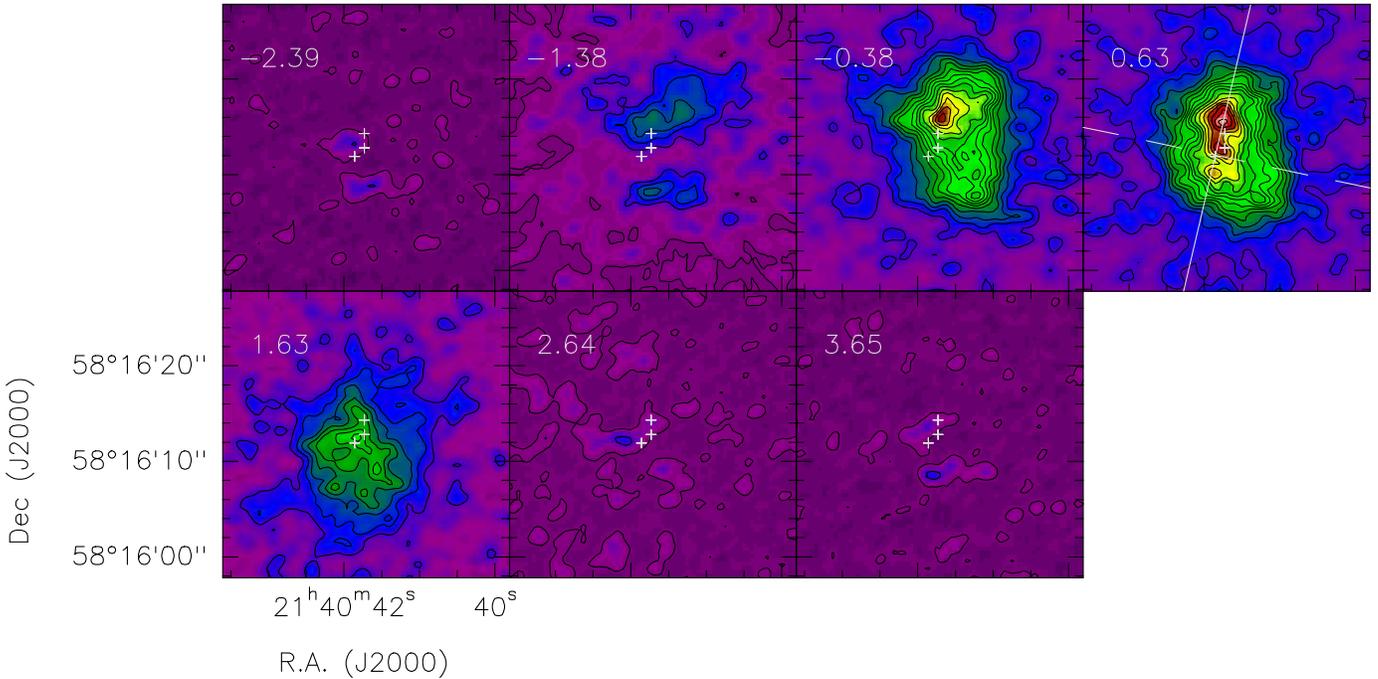}
\vspace{10.5cm}
      \caption{Spectral intensity maps of the N$_2$H$^+$ 1$\rightarrow$0 F=2$\rightarrow$1 line as resulting by combining the
PdBI and 30m data. Each panel is marked with the central velocity. Channel width is $\sim$ 1.0 km/s. Contours are 5 mJy/beam to
380 mJy/beam by steps of 20 mJy/beam.  In the panel labeled
$``$0.63$"$ we have indicated the outflow direction (long dashed line) and the direction of the $``$waist$"$ of gas and dust (solid line). }
         \label{Fig 6}
 \end{figure*}

\section{The CH$_3$CN/N$_2$H$^+$ abundance ratio: a chemical diagnostic}

The different morphologies of CH$_3$CN and N$_2$H$^+$ can be at least qualitatively understood considering that CH$_3$CN is mainly formed on the surface of dust grains (e.g. Bisschop et al. 2007; Garrod et al. 2008), whereas only gas phase processes are responsible for the formation and destruction of N$_2$H$^+$ (Aikawa et al. 2005). On the one hand, the N$_2$H$^+$ abundance increases in cold and dense regions, where CO molecules freeze-out onto dust grains.  On the other hand, the largest fractional abundance of CH$_3$CN is observed toward warm regions (in particular hot cores), where the dust temperature becomes large enough ($\sim$90~K) to allow mantle evaporation. 

To see how the dust temperature affects the CH$_3$CN/N$_2$H$^+$ abundance ratio, we consider  a simple chemical model of a uniform cloud at temperatures between 5 and 100 K. The chemical processes are the same as those described in detail by Caselli et al. (2008): gas phase formation and destruction of N$_2$H$^+$, HCO$^+$, H$_3^+$ and deuterated counterparts,  time-dependent freeze-out of CO and N$_2$, thermal and non-thermal desorption due to cosmic-ray impulsive heating, dust grains with a MRN (Mathis et al. 1977) size distribution. More details about the model can also be found in Emprechtinger et al. (2009). 

For the CH$_3$CN chemistry, we used the results of the comprehensive Garrod et al. (2008) model M, shown in their Fig. 5.  Here, the CH$_3$CN abundance is plotted as a function of time and temperature and it shows two main plateau: one at 40$<$T$<$90 K, where X(CH$_3$CN)$\sim$10$^{-10}$, mainly due to HCN evaporation followed by gas phase processes, and one at T$>$90 K, where X(CH$_3$CN) reaches its peak abundance of 10$^{-8}$ due to direct evaporation from grain mantles. At lower temperatures, X(CH$_3$CN)$<$10$^{-13}$.  To simulate this trend, our model assumes that a fraction (X(CH$_3$CN)=10$^{-10}$) of surface CH$_3$CN has a  binding energy of 1900 K (close to that of CO), whereas the majority has a binding energy of 4500~K (close to that of H$_2$O). 

Figure 10 shows the fractional abundances of CO, N$_2$H$^+$ and CH$_3$CN as a function of gas and dust temperature and for three different values of the gas density: n(H$_2$)=5$\times$10$^4$, 5$\times$10$^5$ and 5$\times$10$^6$ cm$^{-3}$. The gas and dust temperatures are assumed to be the same, given that they are coupled at densities larger than a few 10$^4$ cm$^{-3}$ (Goldsmith 2001).
We assume that the fraction of CH$_3$CN on the surface of dust grains does not change with density. Surface CO completely evaporates at T$>$20 K. As expected, the N$_2$H$^+$ abundance shows a peak at the minimum of the CO abundance (CO is one of the main destroyers of N$_2$H$^+$, together with electrons and negatively charged dust grains). Moreover, being a molecular ion, N$_2$H$^+$ is sensitive to the volume density, given that the electron fraction approximately varies as n(H$_2$)$^{-0.5}$.
The [CH$_3$CN]/[N$_2$H$^+$]  ratio as a function of temperature (and for the three different density values considered) is also plotted in Fig. 10. The model assumes steady state.

The [CH$_3$CN]/[N$_2$H$^+$] ratio varies by 5 orders of magnitude when the temperature increases from 20 to $>$100 K. This
large variation makes this ratio an excellent chemical thermometer in this range of gas kinetic
temperatures. Variation in the molecular hydrogen density can also affect this ratio but to a
lesser extent. Since the N$_2$H$^+$ fractional abundance is dependent on the molecular hydrogen density, the [CH$_3$CN]/[N$_2$H$^+$] ratio together with X(N$_2$H$^+$) are excellent tracers of the temperature and the density of the emitting gas.

\begin{figure*}
\includegraphics{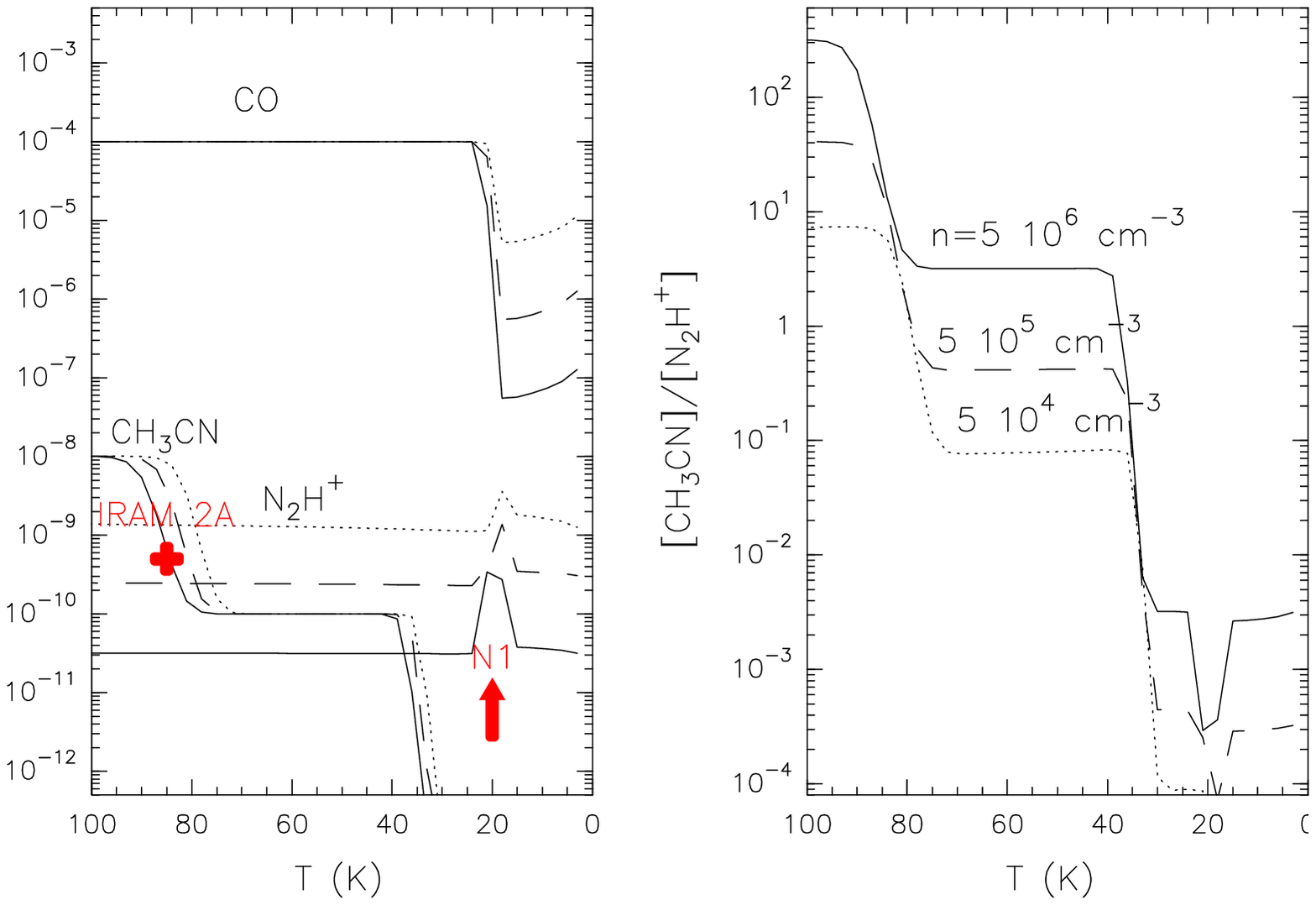}
\vspace{12.0cm}
\caption{
Fractional abundances of CH$_3$CN, N$_2$H$^+$ and CO ({\it left}) and CH$_3$CN/N$_2$H$^+$ abundance ratio({\it right})  as a function of the gas temperature for three different values of the molecular hydrogen density, 5$\times$10$^4$ cm$^{-3}$ (short dashed lines), 5$\times$10$^5$ cm$^{-3}$ (long dashed lines) and 5$\times$10$^6$ cm$^{-3}$ (solid lines).}
\label{rmon_fit}
\end{figure*}

\subsection {Molecular cores in IC~1396~N}

On basis of the results of our model, we are going to use the chemistry as a diagnostic for the physical conditions of the cores
in IC~1396~N. Towards IRAM 2A we have estimated X(CH$_3$CN)$\sim$5~10$^{-10}$ and X(N$_2$H$^+$)$<$1.2 10$^{-10}$. 
The [CH$_3$CN]/[N$_2$H$^+$] ratio and the derived X(CH$_3$CN) are consistent with our chemical model for gas temperatures $\sim$80~K, in
agreement with our estimate of the gas temperature from the CH$_3$CN rotational diagram. The upper limit to the N$_2$H$^+$ abundance derived towards this core suggests that the molecular hydrogen density is $>$5$\times$10$^5$ cm$^{-3}$, which is also consistent with the masses and sizes derived from the continuum emission. The CH$_3$CN abundance measured towards IRAM~2A is similar to that found in hot corinos (Bottinelli et al. 2004, 2007) but 1 to 2 orders of magnitude lower than those expected in high-mass cores (Nummelin et al. 2000; Wilner, Wright \& Plambeck 1994). Fuente et al. (2005b) derived a CH$_3$CN abundance of $\sim$ 7~10$^{-9}$ towards the IM core NGC~7129--FIRS 2 which is more similar to the values found in high mass stars. NGC 7129--FIRS 2 and IC 1396 N are IM protostars with similar luminosities. The difference in the CH$_3$CN abundance of NGC 7129--FIRS 2 and IC~1396~N  could be due to the fact that IC~1396~N is a cluster of low-mass and Herbig Ae stars. In contrast, Fuente et al. (2005b) did not detect clustering in NGC~7129--FIRS~2 down to a spatial resolution of $\sim$1000~AU, which suggests that NGC 7129--FIRS~2 could be the precursor of a more massive Be star. On the other hand, IC~1396~N is a more evolved object than NGC~7129--FIRS~2. While IC~1396~N is considered a borderline Class 0/I object), NGC~7129--FIRS~2 is one of the youngest Class 0 objects known thus far (Fuente et al. 2001, 2005a).
Other possibility is that the low abundance of CH$_3$CN is due to the fact that the recently formed IM star has already formed a small (undetectable with our angular resolution) photodissociation region (PDR). A PDR of Av$\sim$1-3~mag, at a density of n(H$_2$)=5~10$^5$cm$^{-3}$, would have a size of a few 100~AU.
The CH$_3$CN 5$_k$$\rightarrow$4$_k$ and N$_2$H$^+$ 1$\rightarrow$0 lines have not been detected towards BIMA~3. The non-detection of this core in N$_2$H$^+$ is consistent with very high density (n(H$_2$)$>$5$\times$10$^5$ cm$^{-3}$) and hot ($\sim$100~K) gas.

In addition to the IRAM 2A and BIMA 3 cores, we have detected filaments and clumps in the N$_2$H$^+$ 1$\rightarrow$0 line emission that remain undetected in the other tracers. In Table 2 we show the derived N$_2$H$^+$ column densities towards 5 selected positions (labeled 'N' in Fig. 4 and Table 2) assuming gas temperatures of 100, 50 and 20~K. These column densities are in fact lower limits to the real ones since the emission close to the ambient velocities could be underestimated because of the PdBI filtering. We also show lower limits to the N$_2$H$^+$ abundance and upper limits to the [CH$_3$CN]/[N$_2$H$^+$] ratio. Comparing the [CH$_3$CN]/[N$_2$H$^+$] ratio with our chemical calculations (see Fig. 10), we estimate an upper limit of 50~K for the actual gas temperature. However, assuming T$\sim$50~K, the derived N$_2$H$^+$ abundance would be 1 order of magnitude higher than that expected from our chemical model for reasonable values of the molecular hydrogen density ($\sim$5$\times$10$^5$ cm$^{-3}$). Thus, we conclude that these clumps have very likely gas temperatures of around 20~K or less. 
Note that our chemical model predicts a  peak in N$_2$H$^+$ abundance around T=17~K. The UV radiation from the star and possible low velocity shocks are heating the walls of the cavity excavated by the bipolar outflow. In addition to the density and column density enhancements expected in the walls of the cavity, the detection of clumps N1 to N4
might be favored by the warmer gas.

Summarizing, we have detected two chemically different regions in IC1396~N: (i) the warm cores IRAM 2A and BIMA 3, detected in continuum emission and with temperatures around 100~K and (ii) the filaments and the clumps N1 to N4 located along the walls of the bipolar cavity excavated by the outflow that are only detected in the N$_2$H$^+$ 1$\rightarrow$0 line and have very likely temperatures of around 20~K. N5 could be a colder and denser one. The [CH$_3$CN]/[N$_2$H$^+$] ratio is a good chemical diagnostic to discern the temperature of the dense cores.
\begin{table*}
\caption{Molecular cores in IC~1396~N}             
\label{table:1}      
\begin{tabular}{l c c c c c c c c c c}     
\hline\hline       
\multicolumn{1} {c}{Source}  & 
\multicolumn{1}{c}{RA(2000)} &  
\multicolumn{1}{c}{Dec(2000)} &
\multicolumn{1}{c}{Size} &  
\multicolumn{1}{c}{T} &  
\multicolumn{1}{c}{Mass} & 
\multicolumn{1}{c}{N(H$_2$)} &    
\multicolumn{1}{c}{X(N$_2$H$^+$)} &  
\multicolumn{1}{c}{X(CH$_3$CN)} &
\multicolumn{1}{c}{[CH$_3$CN]/[N$_2$H$^+$]} &  
\multicolumn{1}{c}{$\Delta$ v} \\
\multicolumn{1} {c}{}  &  
\multicolumn{1}{c}{} & 
\multicolumn{1} {c}{}  &   
\multicolumn{1}{c}{} &
\multicolumn{1}{c}{} & 
\multicolumn{1}{c}{(M$_\odot$)} & 
\multicolumn{1}{c}{} & 
\multicolumn{1}{c}{} & 
\multicolumn{1}{c}{} & 
\multicolumn{1}{c}{} & 
\multicolumn{1}{c}{(km/s)}  \\
\hline
IRAM 2A$^*$     & 21:40:41.86  &  58:16:11.9  &  0.4$"$$\times$0.2$"$ & {\bf 100} & {\bf 0.06} &
{\bf 3.1~10$^{24}$} &  {\bf $<$1.4~10$^{-10}$}  &  {\bf $\sim$5.0~10$^{-10}$} & $>$4 & $\sim$2  \\ \\

BIMA3$^*$       & 21:40:42.84  &  58:16:01.4  & 0.8$"$$\times$0.4$"$ & {\bf 100}  & {\bf 0.05} &
{\bf 6.5~10$^{23}$} &  {\bf $<$1.6~10$^{-10}$}  &  {\bf $<$2.3~10$^{-9}$} &  &  $\sim$2 \\

            &               &             &      & 50  & 0.11 &      1.4~10$^{24}$    &  $<$4.1~10$^{-11}$       & $<$1.2~10$^{-9}$ & & \\ 
            &               &             &      & 20  & 0.32 &      4.1~10$^{24}$    &  $<$6.2~10$^{-12}$       & $<$6.1~10$^{-10}$ & &     \\ \\
N1$^a$       &  21:40:41.76 &  58:16:15.57 & 1.4$"$$\times$1.2$"$ &  100  &  $<$0.03  &  $<$6.8 10$^{22}$   & $>$1.8 10$^{-9}$  &   & $<$2 & 2.4 \\
         &              &                  &                      &  50   &  $<$0.06  &  $<$1.4 10$^{23}$   & $>$4.7 10$^{-10}$ &   &  $<$5 &    \\
         &              &                  &                      &  {\bf 20} &  {\bf $<$0.15} &  
{\bf $<$3.7 10$^{23}$}   & {\bf $>$8.0 10$^{-11}$}  &   &  $<$16 &    \\ \\
N2$^a$      &  21:40:41.03 &  58:16:18.66 &  1.4$"$$\times$1.2$"$ &  100  & $<$0.03  & $<$6.8 10$^{22}$   & $>$1.1 10$^{-9}$  &   &  $<$5  &   2.1 \\
         &              &                 &                       &  50   & $<$0.06  & $<$1.4 10$^{23}$   & $>$3.0 10$^{-10}$ &   &  $<$8 &    \\
         &              &                 &                       &  {\bf 20}  &  {\bf $<$0.15} & {\bf $<$3.7 10$^{23}$}   &  {\bf $>$5.1 10$^{-11}$}  &  &   $<$40 &    \\ \\
N3$^a$      &  21:40:41.83 &  58:16:08.22 &  1.4$"$$\times$1.2$"$ &  100  & $<$0.03  & $<$6.8 10$^{22}$   & $>$1.6 10$^{-9}$  &   & $<$2    &  1.9 \\
         &              &              &                          &  50   & $<$0.06  & $<$1.4 10$^{23}$   & $>$4.2 10$^{-10}$ &   &  $<$5 &    \\
         &              &              &                        &  {\bf 20}  & {\bf $<$0.15} & {\bf $<$3.7 10$^{23}$}   &  {\bf $>$7.2 10$^{-11}$}  &   &   $<$17 &   \\ \\
N4$^a$      &  21:40:41.22 &  58:16:07.26 & 1.4$"$$\times$1.2$"$ &  100  & $<$0.03  & $<$6.8 10$^{22}$   & $>$1.2 10$^{-9}$  &   &  $<$3  & 2.4 \\
         &              &              &                         &  50   & $<$0.06  & $<$1.4 10$^{23}$  & $>$3.1 10$^{-10}$  &   &   $<$7  & \\
         &              &              &                         &  {\bf 20}  &  {\bf $<$0.15} & {\bf $<$3.7 10$^{23}$}   &  {\bf $>$5.3 10$^{-11}$}  &   &  $<$36 &   \\ \\
N5$^a$      &  21:40:41.81 &  58:15:56.82 & 1.4$"$$\times$1.2$"$ &  100  &  $<$0.03  & $<$6.8 10$^{22}$   & $>$1.3 10$^{-9}$  &    &   $<$3 & 2.1  \\
         &              &              &                         &  50   &  $<$0.06  &  $<$1.4 10$^{23}$   & $>$3.4 10$^{-10}$ &    &   $<$7  &   \\
         &              &              &                        &  {\bf 20}  &   {\bf $<$0.15} & {\bf $<$3.7 10$^{23}$}   &  {\bf $>$5.8 10$^{-11}$}  &   &   $<$20 &  \\
\\  \hline \hline
\end{tabular}

\noindent
$^*$ Sizes, masses and N(H$_2$) from Neri et al. (2007)\\
\noindent
$^a$ The size is assumed to be equal to the beam size. The values of N(H$_2$) and the masses have been estimated from the 3$\times$rms (1mJy/beam) of our 3.1mm map and $\kappa$$_{1mm}$=0.01 cm$^2$ g$^{-1}$ and a dust emissivity spectral index of 1.

\end{table*}
%
%
\begin{figure*}
\includegraphics{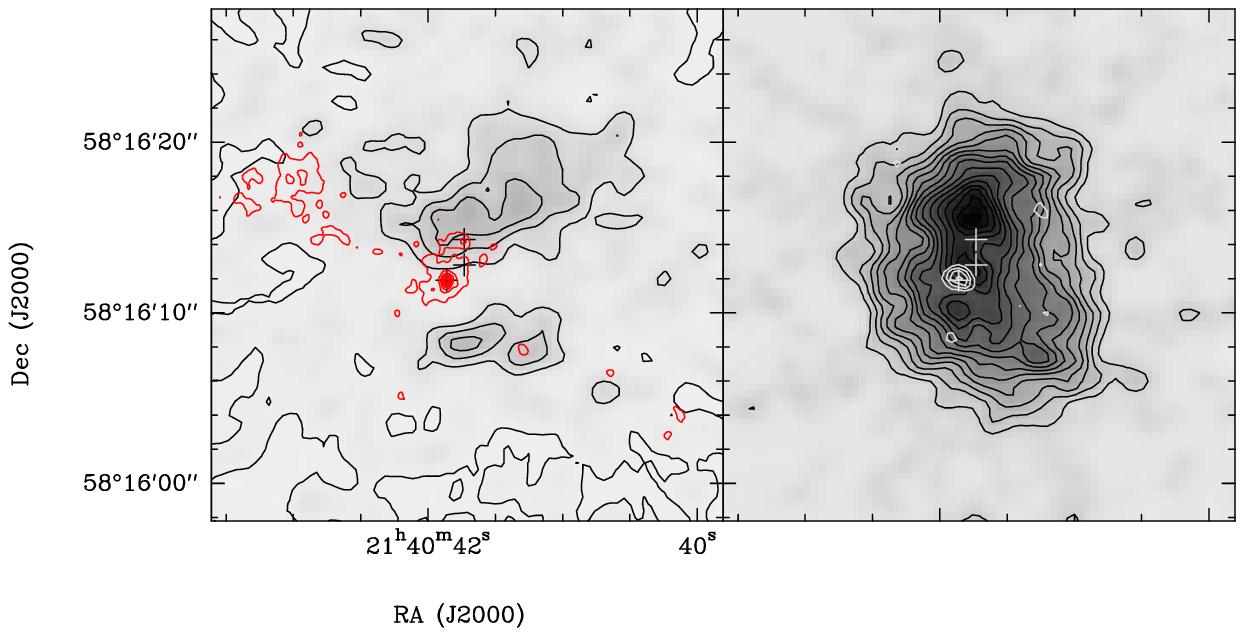}
\vspace{9cm}
      \caption{{\it Left:} spectral map of the N$_2$H$^+$ 1$\rightarrow$0 line emission as resulting from combining our PdBI and
30m data at a velocity of -1.4~km~s$^{-1}$ (grey scale map). This blue-shifted emission delineates the walls of the cavity excavated by the outflow. We have superposed the contours
5 mJy/beam to 35 mJy/beam by steps of 5 mJy/beam of the 1.2mm continuum emission BIMA+PdBI image (dashed red lines) .
{\it Right:} spectral map of the N$_2$H$^+$ 1$\rightarrow$0 line emission as resulting from combining our PdBI and
30m data at the ambient velocity, $\sim$0.63~km~s$^{-1}$ (grey scale map). In solid white contours, we show the integrated intensity PdBI map of the CH$_3$CN 5$_k$$\rightarrow$4$_k$ line. }
         \label{Fig 11}
 \end{figure*}

\section {Overview of the region}
Figure 11 shows an
 sketch of the cometary bright-rimmed globule IC1396~N. The bipolar molecular outflow 
associated with the young intermediate mass protostar has excavated the  molecular
cloud producing a biconical cavity. This cavity has disrupted the cloud in the south-western direction
allowing the high-velocity gas to travel away. The dense gas is located
in an elongated feature, probably an asymmetric toroid, in the direction perpendicular to the outflow. 
The molecular toroid presents strong temperature gradients which produce a chemical differentiation. 

The N$_2$H$^+$ 
emission is very intense in the outer part of the toroid, with the peaks located $\sim$3$"$ North and South from 
the position of the protocluster (see Fig. 9), but absent in the inner region. 
The inner region was modeled by Neri et al. (2007) based on the continuum data. They concluded that the
continuum emission arises in  3 dense cores immersed in a cocoon. The non detection of these cores in N$_2$H$^+$
suggest that all three are warm cores, probably protostars.
Consequently, the previously thought single IM protostar, could actually be a tight proto-cluster. 
Emission of the CH$_3$CN  5$\rightarrow$4 and 13$\rightarrow$12 lines has been detected towards
the most intense core, IRAM 2A (Neri et al. 2007, this Paper). On basis of our CH$_3$CN observations, 
we determine a temperature of 97$\pm$25~K  and a CH$_3$CN abundance of 5$\pm$3~10$^{-10}$ towards IRAM 2A,
similar to that found in hot corinos. 

 Summarizing, the strong dependence of the [CH$_3$CN]/[N$_2$H$^+$] on the 
gas and dust temperature produces a layered structure in the molecular toroid
perpendicular to the bipolar outflow, 
with the CH$_3$CN emission being more intense in the inner region and
N$_2$H$^+$ in the outer part.

\subsection{Source(s) of the molecular outflow(s)}
Three bipolar outflows have been identified in the CO maps: the first one is located around the position of a strong mm continuum BIMA 2 source in the head of the globule, the second one is associated with BIMA~1 (Beltr\'an et al. 2002) and the last one is located in the northern region, outside our interferometric images (Codella et al. 2001).
Single-dish observations of the SiO lines towards the central outflow (that associated with BIMA~2) revealed a highly collimated structure with four clumps of sizes $<$0.1 pc located along the outflow axis. Interferometric images of the CO emission showed, however, that this outflow present a quite complex morphology that Beltr\'an et al. (2002) interpreted as the result of the interaction of the high velocity gas with dense clumps
surrounding the protostar. 

Our highest angular resolution images allow us to learn a bit more about the structure of this complex bipolar outflow. One of the main questions is the driving source of this energetic outflow. The previously thought unique IM protostar BIMA 2, turned to be a cluster of dense cores. The detection of warm CH$_3$CN in IRAM 2A suggest that this is the most massive protostar and could be the driving source of this energetic outflow. This interpretation is also supported by the morphology of the 1mm continuum emission. However the angular resolution of previous interferometric CO observations ($\sim$ 2$"$) does not allow to undoubtedly conclude about that (see Fig. 7). Higher angular resolution observations of CO and/or SiO are required to determine the outflow(s) exciting source(s) 

\section{Conclusions}
We have carried out high-angular resolution (1.4$"$) observations in the continuum at 3.1mm and in the N$_2$H$^+$ 1$\rightarrow$0 and CH$_3$CN 5$_k$$\rightarrow$4$_k$ lines using the Plateau de Bure Interferometer (PdBI). In addition, we have merged the PdBI images with previous BIMA (continuum data at 1.2mm and 3.1mm) and single-dish (N$_2$H$^+$) data to have a comprehensive description of the region. Our results can be summarized as follows:
\begin{itemize}
\item
At large scale, the combination of our data with previous BIMA and 30m data show that the bipolar outflow associated has completely eroded the initial molecular globule. The 1.2mm and 3.1mm continuum emissions are extended along the outflow axis tracing the warm walls of the biconical cavity. 

\item
Most of the molecular gas is located in an elongated feature in the direction perpendicular to the outflow. 
Our results show two types of region in IC1396~N: (i) the cores detected in dust continuum emission, one of which (the most massive) has also been detected in the CH$_3$CN 5$_k$$\rightarrow$4$_k$ line, and (ii) the filaments and clumps located in the molecular toroid, mainly along the walls of the bipolar cavity excavated by the outflow, that are only detected in the N$_2$H$^+$ 1$\rightarrow$0 line. 
This chemical differentiation can be understood in terms of the temperature dependent behavior of the chemistry of N$_2$H$^+$ and CH$_3$CN.  The [CH$_3$CN]/[N$_2$H$^+$] ratio increases by 5 orders of magnitude when the gas temperature increases from 20 to 100~K. 

\item
We have used the [CH$_3$CN]/[N$_2$H$^+$] ratio as a chemical diagnostic to derive the temperature and evolutionary status of the young stellar objects (YSOs). The CH$_3$CN abundance towards IRAM~2A is similar to that found in hot corinos and lower than that expected towards IM and high mass hot cores. This could indicate that IRAM~2A is a low mass or at most Herbig Ae star (IRAM 2A) instead of the precursor of a massive Be star. Alternatively, the low CH$_3$CN abundance could also be the consequence that IRAM 2A is a Class 0/I transition object which has already formed a small photodissociation region (PDR).

\end{itemize}

Our chemical model and observational data prove that the [CH$_3$CN]/[N$_2$H$^+$] ratio is a good tracer of the gas kinetic temperature that is closely related with the spectral type of the star that is being formed. Together with the [CH$_3$CN]/[N$_2$H$^+$] ratio, the fractional abundances of CH$_3$CN and N$_2$H$^+$ can inform about the physical conditions of the gas. Interferometric observations of CH$_3$CN and N$_2$H$^+$ towards IM protostars are, consequently, a valuable tool to have insight into the structure of these young stellar objects.

\begin{acknowledgements}
We are grateful to the IRAM staff in Grenoble and Spain with their great help during the observations and data reduction.
\end{acknowledgements}


\begin{thebibliography}{}

\bibitem[Alonso-Albi et al.(2008)]{alo08} Alonso-Albi, T., 
Fuente, A., Bachiller, R., Neri, R., Planesas, P., 
\& Testi, L.\ 2008, ApJ, 680, 1289 

\bibitem[Alonso-Albi et al.(2009)]{alo09a} Alonso-Albi, T., Fuente, A., Bachiller, R., Neri, R., Planesas, P., Testi, L., Bern{\'e}, O., \& Joblin, C.\ 2009a, A\&A, 497, 117 

\bibitem[Alonso-Albi et al. 2009]{alo09b}
Alonso-Albi et al. 2009b, in preparation

\bibitem[Aikawa et al. 2005]{alo05}
Aikawa, Y., Herbst, E., Roberts, H., \& Caselli, P.\ 2005, ApJ, 620, 330 

\bibitem[Beltr\'an et al. 2002]{bel02}
Beltr{\'a}n, M.~T., Girart, J.~M., Estalella, R., Ho, P.~T.~P., \& Palau, A.\ 2002, ApJ, 573, 246 

\bibitem[Beltr{\'a}n et  al.(2004)]{bel04a} Beltr{\'a}n, M.~T., Gueth, F., Guilloteau, S., \& Dutrey, A.\ 2004a, \aap, 416, 631 

\bibitem[Beltr\'an et al. 2004]{bel04b}
Beltr{\'a}n, M.~T., Girart, J.~M., Estalella, R., \& Ho, P.~T.~P.\ 2004b, A\&A, 426, 941

\bibitem[Beltr\'an et al. 2009]{bel09}
Beltr\'an, M.T., Massi, F., L\'opez, Girart, J.M., Estalella, R. 2009, A\&A, in
press

\bibitem[Bisschop et al. 2007]{bis07}
Bisschop, S.~E., J{\o}rgensen, J.~K., van Dishoeck, E.~F., \& de Wachter, E.~B.~M.\ 2007, \aap, 465, 913 

\bibitem[Bottinelli et al.(2004)]{bot04} Bottinelli, S., et 
al.\ 2004, \apjl, 617, L69 

\bibitem[Bottinelli et 
al.(2007)]{bot07} Bottinelli, S., Ceccarelli, C., Williams, J.~P., \& Lefloch, B.\ 2007, \aap, 463, 601 

\bibitem[Caselli et al. 2008]{cas08}
Caselli, P., Vastel, C., Ceccarelli, C., van der Tak, F.~F.~S., Crapsi, A., \& Bacmann, A.\ 2008, \aap, 492, 703 

\bibitem[Codella et al. 2001]{cod01}
Codella, C., Bachiller, R., Nisini et al.  2001, A\&A, 376, 271 

\bibitem[Emprechtinger et al. 2009]{emp09}
Emprechtinger, M., Caselli, P., Volgenau, N.~H., Stutzki, J., \& Wiedner, M.~C.\ 2009, \aap, 493, 89 

\bibitem[Fuente et 
al.(2001)]{fue01} Fuente, A., Neri, R., Mart{\'{\i}}n-Pintado, J., Bachiller, R., Rodr{\'{\i}}guez-Franco, A., \& Palla, F.\ 2001, \aap, 366, 873 

\bibitem[Fuente et al. 2003]{fue03}
Fuente, A., Rodr{\'{\i}}guez-Franco, A., Testi, L., et al.  2003, ApJL, 598, L39 

\bibitem[Fuente et al.(2005)]{fue05a} Fuente, A., Neri, R., \& Caselli, P.\ 2005a, \aap, 444, 481 

\bibitem[Fuente et al.(2005)]{fue05b} Fuente, A., Rizzo, J.~R., Caselli, P., Bachiller, R., \& Henkel, C.\ 2005b, \aap, 433, 535 

\bibitem[Fuente et al. 2006]{fue06} Fuente, A., Alonso-Albi, 
T., Bachiller, et al. 2006, ApJL, 649, L119

\bibitem[Garrod et al. 2008]{gar08}
Garrod, R.~T., Weaver, 
S.~L.~W., \& Herbst, E.\ 2008, \apj, 682, 283 

\bibitem[Goldsmith(2001)]{gol01} Goldsmith, P.~F.\ 2001, 
ApJ, 557, 736 

\bibitem[Mathis et al. 1977]{mat77}
Mathis, J.~S., Rumpl, 
W., \& Nordsieck, K.~H.\ 1977, \apj, 217, 425 

\bibitem[Neri et al. 2007]{ner07}
Neri, R., et al.\ 2007, A\&A, 468, L33 

\bibitem[Nisini et al. 2001]{nis01}
Nisini, B., et al.\ 2001, \aap, 376, 553 

\bibitem[Nummelin et al.(2000)]{num00} Nummelin, A., Bergman, 
P., Hjalmarson, {\AA}., Friberg, P., Irvine, W.~M., Millar, T.~J., Ohishi, 
M., \& Saito, S.\ 2000, ApJS, 128, 213 

\bibitem[Wilner et al.(1994)]{wil94} Wilner, D.~J., Wright, 
M.~C.~H., \& Plambeck, R.~L.\ 1994, ApJ, 422, 642 


\end{thebibliography}
\end{document}